\documentclass[preprint]{aastex}

\shorttitle{$AKARI$ NIR Spectroscopic Observations of Interstellar Ices in NGC~253}
\shortauthors{M. Yamagishi et al.}

\begin{document}

\title{$AKARI$ Near--Infrared Spectroscopic Observations of Interstellar Ices in Edge-on Starburst Galaxy NGC~253}

\author{Mitsuyoshi Yamagishi\altaffilmark{1}, Hidehiro Kaneda\altaffilmark{1}, Daisuke Ishihara\altaffilmark{1}, Shinki Oyabu\altaffilmark{1}, Takashi Onaka\altaffilmark{2}, Takashi Shimonishi\altaffilmark{2}, and Toyoaki Suzuki\altaffilmark{3}}

\email{yamagishi@u.phys.nagoya-u.ac.jp}

\altaffiltext{1}{Graduate School of Science, Nagoya University, Chikusa-ku, Nagoya 464-8602, Japan}
\altaffiltext{2}{Graduate School of Science, The University of Tokyo, Bunkyo-ku, Tokyo 113-0033, Japan}
\altaffiltext{3}{Institute of Space and Astronautical Science, Japan Aerospace Exploration Agency, Sagamihara, Kanagawa 252-5210, Japan}

\begin{abstract}
We present the spatially-resolved near-infrared (2.5--5.0 $\micron$) spectra of the edge-on starburst galaxy NGC~253 obtained with the Infrared Camera onboard $AKARI$.
Near the center of the galaxy, we clearly detect the absorption features of interstellar ices ($\mathrm{H_2O}$: 3.05 $\micron$, $\mathrm{CO_2}$: 4.27 $\micron$, and $\mathrm{XCN}$: 4.62 $\micron$) and the emission of polycyclic aromatic hydrocarbons (PAHs) at 3.29 $\micron$ and hydrogen recombination line Br$\alpha$ at 4.05 $\micron$.
We find that the distributions of the ices differ from those of the PAH and gas.
We calculate the column densities of the ices and derive the abundance ratios of $N(\mathrm{CO_2})/N(\mathrm{H_2O}) = 0.17 \pm 0.05$.
They are similar to those obtained around the massive young stellar objects in our Galaxy ($0.17 \pm 0.03$), although much stronger interstellar radiation field and higher dust temperature are expected near the center of NGC~253.

\end{abstract}

\keywords{galaxies: individual (NGC~253) --- galaxies: ISM --- infrared: galaxies --- ISM: abundances --- ISM: molecules}


\section{Introduction}

The 2.5--5.0 $\micron$ near-infrared (NIR) spectra of the interstellar media in galaxies are dominated by various emission and absorption features.
For example, the absorption of various ice species (solid-state molecules, e.g. $\mathrm{H_2O}$: 3.05 $\micron$, $\mathrm{CO_2}$: 4.27 $\micron$, $\mathrm{XCN}$: 4.62 $\micron$, and $\mathrm{CO}$: 4.67 $\micron$), as well as the emission of polycyclic aromatic hydrocarbons (PAHs) at 3.29 $\micron$ and hydrogen recombination lines such as Br$\alpha$ at 4.05 $\micron$, are included in the NIR regime.
In particular, ices are important to understand interstellar chemistry, since the absorption profiles of ices are known to be sensitive to the chemical composition and the temperature of dust grains (e.g. Pontoppidan et al. 2008; Zasowski et al. 2009).

Ices around young stellar objects (YSOs) in our Galaxy and the Large Magellanic Cloud (LMC) have been studied well until now (e.g. Gerakines et al. 1999; Gibb et al. 2004).
Shimonishi et al. (2008, 2010) showed that the abundance ratios $N(\mathrm{CO_2})/N(\mathrm{H_2O})$ around massive YSOs in the LMC (0.36 $\pm$ 0.09) are significantly higher than those in our Galaxy (0.17 $\pm$ 0.03; Gerakines et al. 1999; Gibb et al. 2004).
Ices are also detected in Galactic quiescent molecular clouds; Whittet et al. (2007) reported that they show the abundance ratios of 0.18 $\pm$ 0.04.
Ices in nearby galaxies, however, have not been studied well; there are only a few reports about the detection of ices.
Sturm et al. (2000) reported the first detection of $\mathrm{H_2O}$ ice absorption in the NIR and mid-infrared (MIR) spectra of NGC~253 and M~82 with the $ISO$ SWS.
Following the detection of the $\mathrm{H_2O}$ ice, the detection of the $\mathrm{CO_2}$, XCN and CO ices was reported in the nucleus of NGC~4945 (Spoon et al. 2000, 2003).
However spatially-resolved study about ices has not been conducted yet except for the L- and M-band study of the circumnuclear $10\arcsec$ region of NGC~4945 by Spoon et al. (2003).

\begin{figure}
\epsscale{0.7}
\plotone{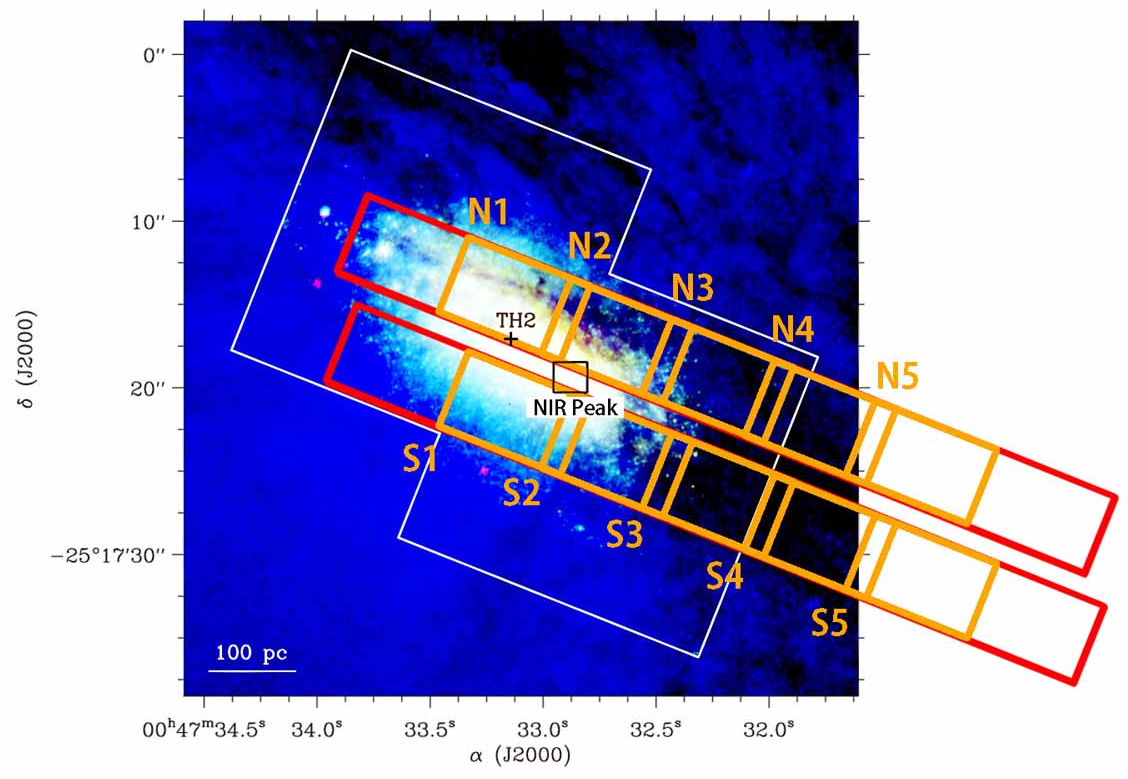}
\plotone{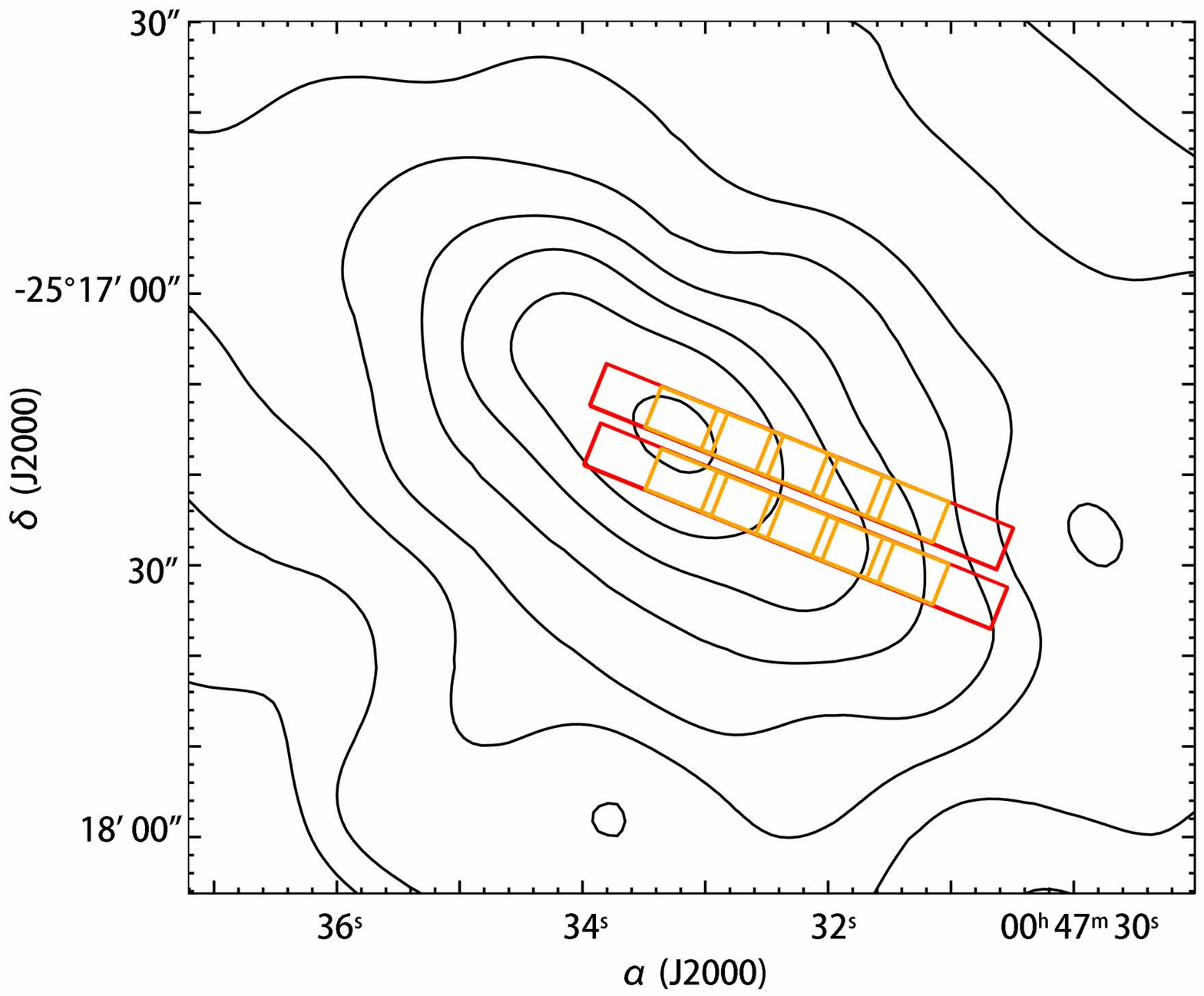}
\figcaption{(top) Slit full apertures (red boxes) and sub-aperture regions (orange boxes with their names) overlaid on the $Hubble$ $Space$ $Telescope$ NIR three color image of NGC~253 taken from Kornei \& McCrady (2009). The color scale of the image is changed from the original one to show low surface brightness area more clearly. Each sub-aperture (N1--5, S1--5) has a size of $5 \arcsec$ $\times$ $ 7.5 \arcsec$ with a separation of 6 $\arcsec$ between each, from which we created the spectra. (bottom) The same apertures overlaid on the integrated $\mathrm{^{12}CO}$ map of NGC~253 (Kuno et al. 2007). The contours are drawn at logarithmically-spaced 8 levels from 95\% to 5\% of the peak intensity (1.08 $\times$ $10^3$ $\mathrm{K~km~s^{-1}}$). \label{region}}
\end{figure}

NGC~253 is a well-studied starburst galaxy at a distance of 3.5 Mpc (Rekola et al. 2005), which has a large inclination angle ($\sim 80^\circ$).
Due to the high inclination angle, we can gain high column densities along the line of sight.
Hence, it is relatively easy to detect various absorption features, if any, from NGC~253.
The kinematic center of NGC~253 is a compact radio source at a wavelength of 2 cm, TH2, while the peak of NIR emission is spatially separated from the TH2 by $4\arcsec$ (see Fig. \ref{region}).
The NIR peak is thought to be a young super star cluster (Keto et al. 1999, Kornei \& McCrady 2009).
In Fig.\ref{region}, prominent dust lanes are visible on the north and the south-west side of the NIR peak.
Kuno et al. (2007) presented the integrated $\mathrm{^{12}CO}$ map of NGC~253 with the beam size of $15\arcsec$.
In the CO map (Fig.\ref{region}), there is no apparent structure corresponding to the NIR dust lane.
The central activity of the galaxy is known to be strong enough to produce prominent X-ray (Dahlem et al. 1998) and H$\alpha$ (Hoopes et al. 1996) as well as large-scale HI plumes (Boomsma et al. 2005).
Moreover, $AKARI$ clearly detected far-infrared dust outflow from the galactic disk (Kaneda et al. 2009b), Tacconi-Garman et al. (2005) showed the distribution of PAH 3.3 $\micron$ emission for the central region of NGC~253 by using the narrow-band images with the VLT.

In this letter, we present the NIR (2.5--5.0 $\micron$) spectra of NGC~253 obtained with the Infrared Camera (IRC; Onaka et al. 2007) on board the $AKARI$ satellite (Murakami et al. 2007).
The spectra clearly show the absorption features of the $\mathrm{H_2O}$ and $\mathrm{CO_2}$ ices.
Based upon the spectra, we discuss the interstellar chemical condition in NGC~253.


\section{Observations and Data Reduction}

The NIR spectroscopic observations were performed as part of the $AKARI$ mission program "ISM in our Galaxy and Nearby galaxies" (ISMGN; Kaneda et al. 2009a) in the $AKARI$ post-helium phase (phase 3).
The observations were carried out on December 21 2009.
To obtain 2.5--5.0 $\micron$ spectra, we used a grism spectroscopic mode (R $\sim$ 120) with the slit of $5 \arcsec \times 48 \arcsec$ for its width and length, respectively (Ohyama et al. 2007).
Figure \ref{region} shows the slit positions  of the observations and the regions from which we created the spectra.
We observed two regions in NGC~253, the north and south sides of the NIR peak (Observation ID: 1422187, 1422196).
To avoid saturation effects, each region was selected not to cover the NIR peak.
We observed each region two times to improve data quality.

The basic spectral analysis was performed by using the standard IDL pipeline prepared for reducing phase 3 data with a newly calibrated spectral response curve\footnote{http://www.ir.isas.jaxa.jp/ASTRO-F/Observation/}.
In addition to the basic pipeline process, we applied the following custom procedures to improve S/Ns for each spectrum: 
before creating a spectrum, we removed hot pixels from the three array images, where pixel intensities are replaced by the median values of contiguous 8 pixels, and then we obtained three spectra for the same region by integrating pixel intensities over the spatial scale of $7.5 \arcsec$ along the direction of the slit length.
Next, we combined the two spectra by calculating a median value of 6 pixels, where 3 pixels in the direction of wavelength per spectrum were considered for the calculation.
Standard deviations were then adopted as flux errors.
Finally, we applied smoothing with a boxcar kernel of 3 pixels ($\sim 0.03$ $\micron$) in the direction of wavelength.
We neglected the background of each spectrum since signals in a region $5 \arcmin$ away from the center of NGC~253 are about a hundred times smaller than those of the center.


\section{Result}

\begin{figure}
\epsscale{1.0}
\plotone{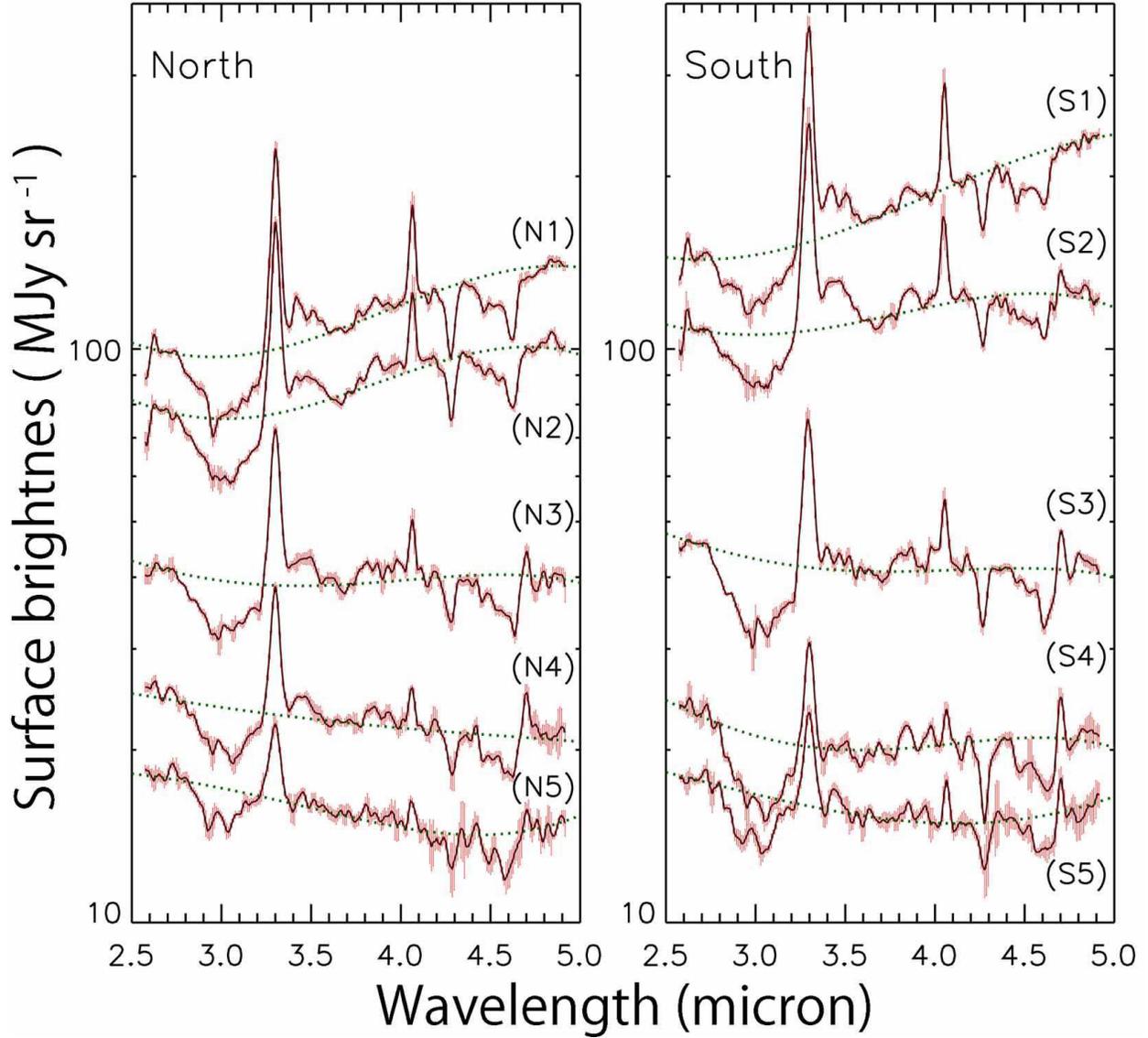}
\figcaption{Spectra obtained for NGC~253 with the names of the regions indicated in Fig. \ref{region}. The left and right panel show the spectra observed in the north and the south region, respectively. The green dotted lines indicate the best-fit continuum curves that we use in deriving the optical depths of the absorption features (see text). \label{spectra}}
\end{figure}

The obtained spectra are shown in Fig  \ref{spectra}.
The surface brightness of the spectra is different from region to region; the S1 and N1 spectra show the highest surface brightness for each slit aperture, which monotonically decreases toward the N5 and S5 spectra.
The slopes of the spectra also change from the N1 and S1 to the N5 and S5 spectra.
Several strong features are detected in the spectra; PAH emission at 3.3 $\micron$, hydrogen recombination line Br$\alpha$ at 4.05 $\micron$, and the absorption of ices.
The absorption features of the $\mathrm{H_2O}$ ice centered at 3.05 $\micron$ and the $\mathrm{CO_2}$ ice at 4.27 $\micron$ are detected in all the spectra.
Some spectra also show the absorption feature of $\mathrm{XCN}$ ice at 4.62 $\micron$ and the pure rotational line of molecular hydrogen $\mathrm{H_2S(9)}$ at 4.69 $\micron$.
With $ISO$, Sturm et al. (2000) reported only the detection of the $\mathrm{H_2O}$ ice, and hence this is the first detection of the $\mathrm{CO_2}$ and XCN ices in NGC~253.
The PAH emission at 3.3 $\micron$ is also detected in all the spectra.
The spatial distribution of the PAH 3.3 $\micron$ emission to the south-west direction from the NIR peak is at least 2.5 times wider than that shown in Tacconi-Garman et al. (2004) owing to high sensitivity in the space observations.

\begin{figure}
\epsscale{0.6}
\plotone{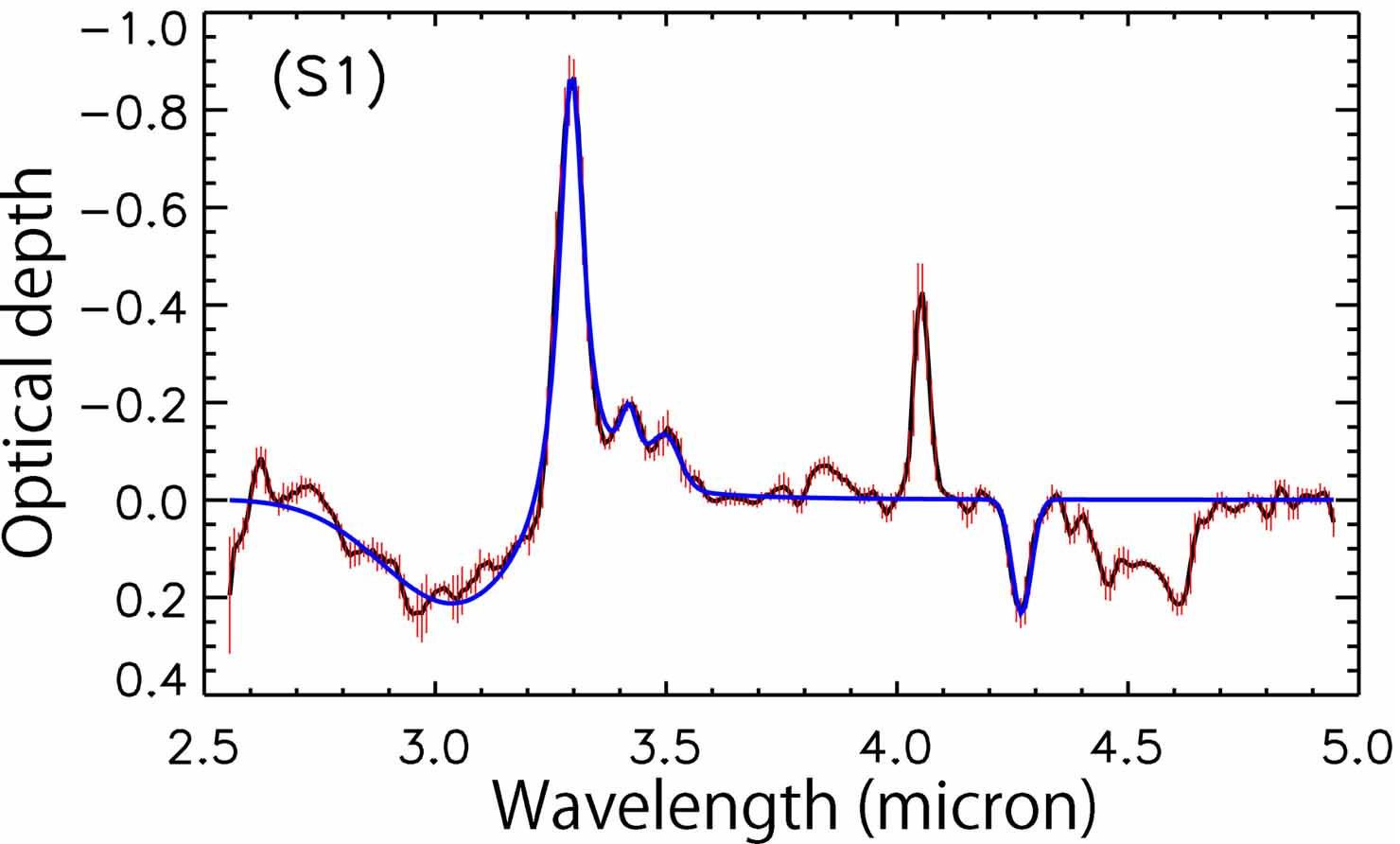}
\plotone{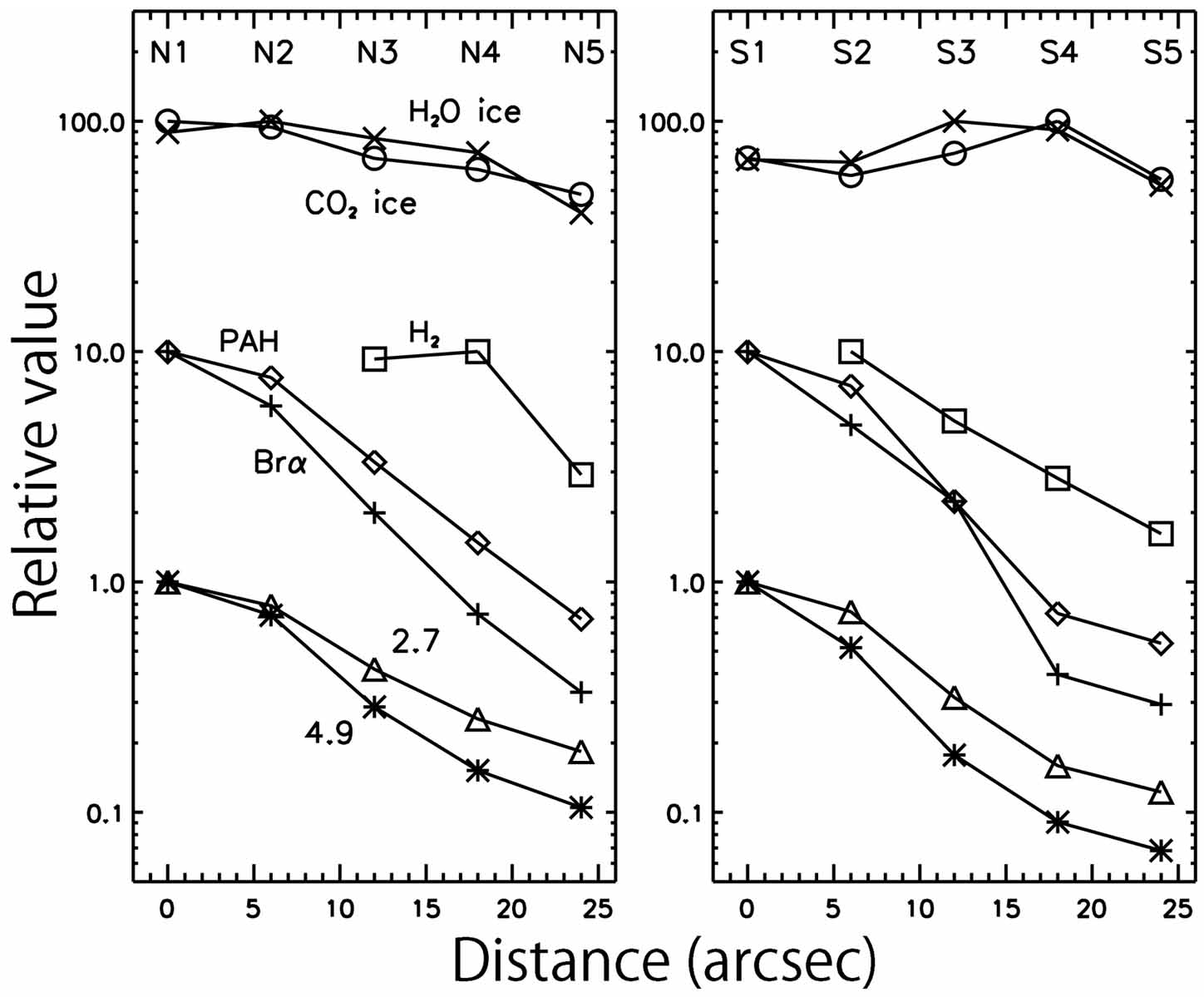}
\figcaption{(top) Example of the optical depth spectrum and the spectral fitting for the S1 region. (bottom) Changes of the ice column densities as well as the intensities of the PAH feature, gas lines, and continuum emissions with the distance from N1 for the north region (left) and S1 for the south region (right). The cross, circle, diamond, plus, square, triangle, and asterisk symbols indicate the column densities of the $\mathrm{H_2O}$ ice (peak values of $5.8 \times 10^{17} \mathrm{cm^{-2}}$ for the north side and $7.4 \times 10^{17} \mathrm{cm^{-2}}$ for the south side) and $\mathrm{CO_2}$ ice ($1.3 \times 10^{17} \mathrm{cm^{-2}}$ and $1.0 \times 10^{17} \mathrm{cm^{-2}}$), the integrated intensities of the PAH 3.3 $\micron$ ($2.5 \times 10^{-5} \mathrm{erg~s^{-1} cm^{-2} sr^{-1}}$ and $4.0 \times 10^{-5} \mathrm{erg~s^{-1} cm^{-2} sr^{-1}}$), Br$\alpha$ ($3.9 \times 10^{-6} \mathrm{erg~s^{-1} cm^{-2} sr^{-1}}$ and $5.6 \times 10^{-6} \mathrm{erg~s^{-1} cm^{-2} sr^{-1}}$), and $\mathrm{H_2 S(9)}$ ($1.6 \times 10^{-7} \mathrm{erg~s^{-1} cm^{-2} sr^{-1}}$ and $4.9 \times 10^{-7} \mathrm{erg~s^{-1} cm^{-2} sr^{-1}}$), and the surface brightness at 2.7 $\micron$ (98 $\mathrm{MJy~sr^{-1}}$ and 145 $\mathrm{MJy ~sr^{-1}}$) and 4.9 $\micron$ (140 $\mathrm{MJy~sr^{-1}}$ and 235 $\mathrm{MJy~sr^{-1}}$), respectively. Each profile is normalized by its peak value. The column densities of the ices and the integrated intensities of the line and feature emissions are multiplied by 100 and 10, respectively for display purpose. \label{tau}}
\end{figure}

To obtain continuum spectra, we fit the continuum regions at 2.65--2.70 $\micron$, 3.60--3.70 $\micron$, 4.10--4.15 $\micron$, 4.35--4.45 $\micron$, and 4.85--4.95 $\micron$ by a fourth order polynomial.
The best-fit continuum curve for each spectrum is shown in Fig  \ref{spectra}.
We divide the original spectra by the continuum spectra to derive the optical depth spectra (Fig \ref{tau}).

To calculate the column densities of the $\mathrm{H_2O}$ ice, we fit a Gaussian profile to the optical depth spectra (Fig \ref{tau}).
Since the present spectra resolve the absorption feature of the $\mathrm{H_2O}$ ice, we can measure a true optical depth.
However we have to consider the contribution of the PAH emission at 3.3 $\micron$ and its sub-features at 3.4 and 3.5 $\micron$ to the absorption of the $\mathrm{H_2O}$ ice.
Hence Lorentzian profiles are included in the model fitting for the 3.3 and 3.4 $\micron$ features and a Gaussian for the 3.5 $\micron$ feature to fit the range of 2.65--3.65 $\micron$.
We first fitted the S1 spectrum, determined the widths of the Gaussian and Lorentzian profiles, and then applied the same widths to the other spectra.
In the spectral fitting, the centers of Gaussian and Lorentzian profiles are fixed at 3.05, 3.29, 3.42, and 3.50 $\micron$ for the $\mathrm{H_2O}$ ice, 3.3, 3.4, and 3.5 $\micron$ features, respectively.
The result of fitting to one of the optical depth spectra is shown in Fig \ref{tau}.
The derived optical depths of the $\mathrm{H_2O}$ ice in the N1 and the S1 region ($\tau = 0.23 \pm 0.2$ for both regions) are consistent with that previously measured by $ISO$ ($\tau \sim 0.25$; Sturm et al. 2000) within the errors; the S1, S2, N1, and N2 regions overlap with the slit aperture of Sturm et al. (2000).
We derive the column density, $N$, from the equation
\begin{equation}
N = \int \tau d\nu/A,
\end{equation}
where $A$, $\tau$, and $\nu$ are the band strength of each ice feature measured in a laboratory, an optical depth, and a wavenumber, respectively.
The band strength of $2.0 \times 10^{18}$ $\mathrm{cm~molecule^{-1}}$ is used for the $\mathrm{H_2O}$ ice (Gerakines et al. 1995).

On the other hand, the present spectra cannot resolve the absorption feature of the $\mathrm{CO_2}$ ice.
We, however, applied the method of integrating the optical depth spectra rather than a curve-of-growth method since the equivalent width of $\mathrm{CO_2}$ ice absorption is very small ($\sim$ 0.01 $\micron$).
Hence we use a Gaussian profile to fit each optical depth spectrum of the $\mathrm{CO_2}$ ice and used the fitting range of 4.20--4.35 $\micron$.
In the above equation, we use the band strength of $7.6 \times 10^{17}$ $\mathrm{cm~molecule^{-1}}$ (Gerakines et al. 1995).
The systematic error of each column density is estimated to be 15 \% for the $\mathrm{H_2O}$ ice and 10 \% for the $\mathrm{CO_2}$ ice.
The errors are evaluated by changing the above-defined continuum regions with small shifts of $\pm 0.05$ $\micron$ for both ices.

The derived column densities of the $\mathrm{H_2O}$ and $\mathrm{CO_2}$ ice, the integrated line intensities of PAH 3.3 $\micron$, Br$\alpha$, and $\mathrm{H_2S(9)}$, and the surface brightness at 2.7 $\micron$ and 4.9 $\micron$ are shown in Fig \ref{tau}.
The surface brightness at 2.7 $\micron$ and 4.9 $\micron$ is the median values over the wavelength ranges of 2.65 to 2.75 $\micron$ and 4.85 to 4.95 $\micron$, respectively.
In Fig \ref{tau}, the spatial profiles of the ices are different from the other features; the integrated line intensity and the surface brightness have peaks at the N1 and S1 region and decrease rapidly to the N5 and S5 region except the integrated intensity of the $\mathrm{H_2S(9)}$, while the column densities of the ices show much smaller changes from region to region.
In addition, on the south side, the ices show peaks in the S3 and S4 regions far from the NIR peak.
These profiles suggest that the absorbers responsible for the ice features are more widely distributed than the line and the continuum emitters.
The profiles of the $\mathrm{H_2O}$ and $\mathrm{CO_2}$ ices are similar to each other, suggesting a good correlation between the column densities of the $\mathrm{H_2O}$ and $\mathrm{CO_2}$ ice, although the phase of the gas dominating the spectral features differs from region to region.

The N1 and N2 spectra show weaker continuum emission than the S1 and S2 spectra due to the presence of the NIR dust lane on the north side of the NIR peak (Fig.\ref{region}).
In the S3 and S4 regions, another prominent dust lane is visible, which presumably contributes to the larger column densities of the ices.
Thus some of the ices responsible for the observed absorption are likely to be associated with the dust lanes, while the others are not.
On the other hand, the distribution of the CO emission does not show clear spatial correspondence with those of the ice absorptions and the dust lanes.
The CO map in Fig.\ref{region} reveals a more centrally-concentrated distribution, which does not have a local maximum around the sub-apertures of S3 and S4, although the beam size of the CO map ($\sim$ $15\arcsec$) is somewhat larger than the spatial scale of the sub-apertures ($\sim$ $7\arcsec$). 
Therefore a majority of the CO molecular clouds do not significantly contribute to the observed absorptions due to the ices.

\begin{figure}
\epsscale{1.0}
\plotone{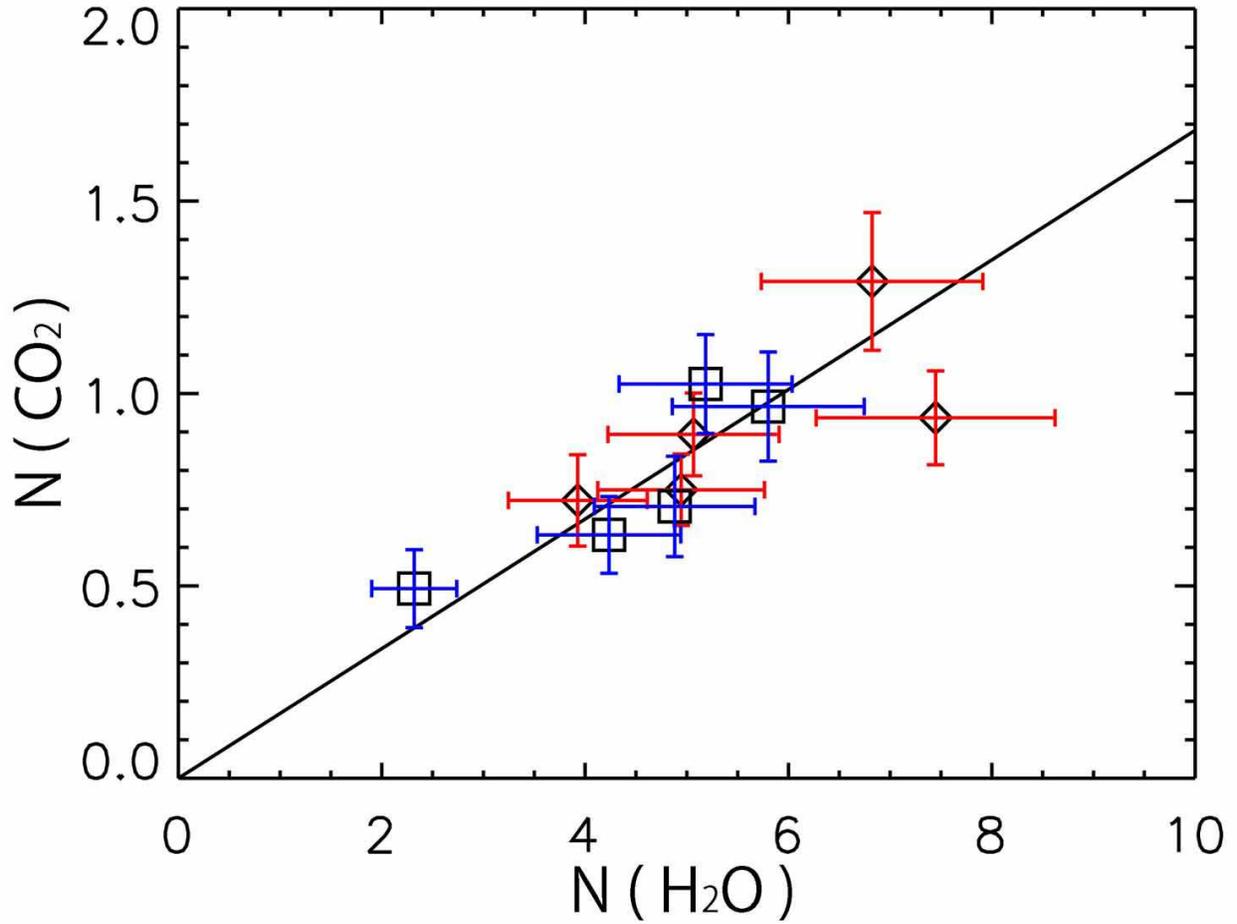}
\figcaption{Column densities of the $\mathrm{CO_2}$ ice plotted against the column densities of the $\mathrm{H_2O}$ ice, both given in units of $10^{17} \mathrm{cm^{-2}}$. The squares with the error bars in blue and the diamonds with the error bars in red represent the results from the north and south regions, respectively. The solid line represents the best-fit linear relation between the $\mathrm{CO_2}$ ice and $\mathrm{H_2O}$ ice column densities. \label{ratio}}
\end{figure}

In Fig \ref{ratio}, we show the correlation plot of the derived column densities of the $\mathrm{H_2O}$ and $\mathrm{CO_2}$ ices, which shows a linear correlation on both sides of the NIR peak.
There is no systematic difference in the relation between the north and south regions.
From the slope of the best-fit line to the data, the averaged $\mathrm{CO_2}$/$\mathrm{H_2O}$ ice ratio is calculated to be $0.17 \pm 0.05$.
The ratio is similar to that obtained from the Galactic massive YSOs ($0.17 \pm 0.03$, Gerakines et al. 1999; Gibb et al. 2004).
In our observation, we detect the superposition of ices in various kinds of clouds present along the line of sight, while the observations of the Galactic YSOs trace the chemical environment of individual star-forming clouds. 
Therefore it is interesting that these observations show similar $\mathrm{CO_2}$/$\mathrm{H_2O}$ ice ratios despite the different situations.

In the above calculation, we assume the following geometry: the continuum emissions are in the background and absorbed by the ices with the covering fraction of 100 \% for each sub-aperture, while the PAH 3.3 $\micron$ emission and its sub-features are distributed in the foreground of the ices.
We also calculate the column densities in the case that the covering fraction of the ice features is 50 \%.
Then, the column densities change to 4.8--16.8 $\times$ $10^{17}$ $\mathrm{cm^{-2}}$ and 1.1--3.1 $\times$ $10^{17}$ $\mathrm{cm^{-2}}$ for the $\mathrm{H_2O}$ and $\mathrm{CO_2}$ ice, respectively, and the $\mathrm{CO_2}$/$\mathrm{H_2O}$ ice ratio of 0.17 $\pm$ 0.05, the same as above, is obtained.
However, if the covering fraction is small, there is a possibility that the observed broad profile of the $\mathrm{H_2O}$ ice feature might be saturated.
We therefore compare the obtained optical depth spectra of the $\mathrm{H_2O}$ ice with those obtained in the laboratory from the Leiden Molecular Astrophysics database (Ehrenfreund et al. 1996).
We use the laboratory profile of the pure $\mathrm{H_2O}$ ice at 10 K.
As a result, we do not find any significant difference between both optical depth spectra.
Therefore, it is unlikely that the observed profiles of the $\mathrm{H_2O}$ ice are saturated.
Moreover, we also calculate the column densities of the $\mathrm{H_2O}$ ice in the case that the PAH 3.3 $\micron$ emission and its sub-features are distributed in the background of the ices.
Then we derive the slightly ($\sim$ 2 \%) smaller column densities of the $\mathrm{H_2O}$ ice than the above calculation.
Thus our results do not significantly depend on the assumed geometry.

On the other hand, Shimonishi et al. (2008, 2010) showed that the $\mathrm{CO_2}$/$\mathrm{H_2O}$ ice abundance ratios of the massive YSOs in the LMC ($0.36 \pm 0.9$) are significantly higher than those in our Galaxy.
Shimonishi et al. (2008, 2010) interpreted that the high ratio is caused by the relative increase of the $\mathrm{CO_2}$ ice possibly due to either strong interstellar ultraviolet (UV) photon irradiation to $\mathrm{H_2O}$-$\mathrm{CO}$ binary ice mixtures (e.g. Watanabe et al. 2007) or relatively high dust temperatures (Bergin, Neufeld, \& Melnick 1999; Ruffle \& Herbst 2001).

In the center of NGC~253, nuclear starburst has occurred (Dudley \& Wynn-Williams 1999), which indicates the existence of strong UV radiation field.
The detection of XCN ice at 4.62 $\micron$ is indicative of strong UV irradiation (Bernstein, Sandford, \& Allamandola 2000; Spoon et al. 2003).
The slopes of the NIR continuum spectra suggest the presence of hot dust.
Therefore our result suggests that intense interstellar UV radiation field and high dust temperatures are not important factors to determine the ice abundance ratio.
Metallicity in NGC~253 is known to be close to a solar value, $Z \sim 1 Z_\odot$ (Webster \& Smith 1983), while $Z \sim 0.3 Z_\odot$ for the LMC (Luck et al. 1998).
Therefore, the interstellar metallicity might be an important chemical condition to affect the ice abundance ratio.


\section{Conclusions}
With $AKARI$, we have performed the NIR (2.5--5.0 $\micron$) spectroscopic observation of the central region of the edge-on starburst galaxy NGC~253.
We clearly detect the absorption features of the $\mathrm{H_2O}$, $\mathrm{CO_2}$, and XCN ices in addition to the PAH 3.3 $\micron$ feature and its sub-features at 3.4--3.5 $\micron$, the hydrogen recombination line Br$\alpha$ at 4.05 $\micron$, the molecular hydrogen pure-rotational line $\mathrm{H_2S(9)}$ at 4.69 $\micron$, and hot dust continuum emission.
We for the first time obtain the spatial variations of the ice absorption features for nearby galaxies.
We find that the ices have different distributions from PAH, ionized gas, molecular gas, and hot dust.
We evaluate the column densities of the $\mathrm{H_2O}$ and $\mathrm{CO_2}$ ices and derive their abundance ratios, $N(\mathrm{CO_2})$/$N(\mathrm{H_2O})$, of $0.17 \pm 0.05$.
The obtained ratios are very close to those observed for massive YSOs in our Galaxy.
However they are significantly lower than those in the LMC where strong UV radiation and high dust temperatures are expected but not as much as in the central region of NGC~253.
Therefore we conclude that intense interstellar UV radiation field and high dust temperatures are not important factors to determine the ice abundance ratio.

\acknowledgments

We would like to thank all the members of the $AKARI$ project for their intensive efforts.
We also express many thanks to the anonymous referee for the useful comments.
This work is based on observations with $AKARI$, a JAXA project with the participation of ESA.

\end{document}